\documentstyle[preprint,floats,tighten,prl,aps]{revtex}
\def \pom {{\scriptscriptstyle \kern -0.1em I \kern -0.25em P}}
\input epsf.tex
\def\desepsf(#1 width #2){\epsfxsize=#2 \epsfbox{#1}}
\begin{document}
\preprint{\vbox{
\hbox{August 2000} }}
\draft
\def\cm{{\cal M}}
\def\as{\alpha_s}
\def\ee{e^+e^-}
\def\qq{q \bar{q}}
\def\lmsb{\Lambda_{\overline{\rm MS}}}
\def\to{\rightarrow}
\def\fb{~{\rm fb}}
\def\pb{~{\rm pb}}
\def\ev{\,{\rm eV}}
\def\kev{\,{\rm KeV}}
\def\mev{\,{\rm MeV}}
\def\gev{\,{\rm GeV}}
\def\GeV{\,{\rm GeV}}
\def\tev{\,{\rm TeV}}
\def\ep{\epsilon}
\def\nn{\nonumber}
\def\pen{\frac{1}{(N+1)(N+2)} }
\def\peng{\frac{1}{(N+2)} }
\newcommand\f[2]{\frac{#1}{#2}}

\title{\large \bf Next-to-next-to-leading logarithmic corrections at small transverse
  momentum in hadronic collisions\footnote{Work supported in part by the EU Fourth Framework Programme `Training and Mobility of Researchers', Network `Quantum Chromodynamics and the Deep Structure of Elementary Particles', contract FMRX-CT98-0194 (DG 12 - MIHT) and by the Swiss National Foundation.}}

\author{ {\bf Daniel\ de Florian} and {\bf Massimiliano\ Grazzini} }

\address{Institute of Theoretical Physics, 
ETH-H\"onggerberg, CH-8093 Z\"urich, Switzerland}

\maketitle

\hspace{1cm}

\begin{abstract}

We study the region of small transverse momenta in $q{\bar q}$- and
$gg$-initiated processes with
no colored particle detected
in the final state. We present the 
universal expression of the ${\cal O}(\as^2)$ logarithmically-enhanced 
contributions up to next-to-next-to-leading logarithmic accuracy.
From there we extract  the coefficients that allow the resummation of the 
large logarithmic contributions.
We find that the coefficient known in the literature as $B^{(2)}$
is process dependent, since it receives a hard contamination from the one
loop correction to the leading order subprocess.
 We present the general result of
 $B^{(2)}$ for both quark and gluon channels.
In particular,
in the case of Higgs production, this result will be relevant to improve the
matching between resummed predictions and fixed order calculations.

\end{abstract} 
\vspace{1cm}

\narrowtext

The process in which a system of non strongly-interacting particles of large
invariant mass $Q^2$ (lepton pairs, gauge boson(s), Higgs boson, and so forth)
 is produced in
hadronic collisions is a well studied subject in perturbative QCD \cite{lhcworkshop}. At
transverse momenta $q_T^2$ of order of $Q^2$ the cross section can be computed
by using the standard QCD-improved parton model. When $q_T$ becomes small the
simple perturbative picture is spoiled. This happens because large logarithmic 
corrections of the form $\log\f{Q^2}{q_T^2}$ arise due to a non complete
cancellation of soft and collinear singularities between real and virtual
contributions. These large logarithmic corrections can be resummed to all
orders by using the Collins-Soper-Sterman (CSS) formalism \cite{CSS}.

We consider the class of inclusive hard scattering processes
\begin{equation}
\label{class}
h_1 h_2\to A_1+A_2\dots A_n+X
\end{equation}
where the collision of the hadrons $h_1$ and $h_2$ produces a system of non 
strongly-interacting
final state particles $A_1\dots A_n$ 
carrying total momentum $Q$ and total transverse momentum $q_T$. 
According to the CSS formula, and neglecting terms which are finite in the limit $q_T\rightarrow 0$, the 
cross-section can be
written as 
\footnote{It is assumed that all other dimensionful invariants are of the same
  order $Q^2$.} :
\begin{eqnarray}
\label{CSS}
\f{d\sigma}{dq_T^2 dQ^2 d\phi}&=& \sum_{a,b,c} \int_0^1 dx_1 \int_0^1 dx_2
\int_0^\infty db \f{b}{2} J_0(bq_T)\,\f{d\sigma_{c\bar{c}}^{(LO)}}{d\phi} \,\delta(Q^2-x_1 x_2 s)\,\nn\\
&\cdot&\left(f_{a/h_1}\otimes C_{ca}\right)\left(x_1,\f{b_0^2}{b^2}\right) \,
\left(f_{b/h_2}\otimes C_{\bar{c}b}\right)\left(x_2,\f{b_0^2}{b^2}\right)
S_c(Q,b)\, 
,
\end{eqnarray}
where $d\phi=dPS(Q\to q_1,q_2,\dots q_n)$ represents the phase space of the
system of non-colored particles, $b_0=2 e^{-\gamma_e}$ and
$\sigma_{c\bar{c}}^{(LO)}$ is the leading-order cross-section (i.e., with no
final state partons and therefore $q_T=0$) for the given 
process ($c,\bar{c}$ can be either $q_f, \bar{q}_{f'}$ or $g,g$).
The function $C_{ab}$ 
in Eq.~(\ref{CSS}) is a process-dependent coefficient
function,
$J_0(b q_T)$ is the Bessel function of first kind and $f_{i/h}$ corresponds to 
the distribution of a parton $i$ in a hadron $h$. The large logarithmic corrections are exponentiated in the
Sudakov form factor
\begin{equation}
\label{sudakov}
S_c(Q,b)=\exp \left\{ -\int_{b_0^2/b^2}^{Q^2} \frac{dq^2}{q^2} 
\left[ A_c(\as(q^2)) \;\ln \frac{Q^2}{q^2} + B_c(\as(q^2)) \right] \right\}\, .
\end{equation}
The functions $A_c$, $B_c$ and $C_{ab}$ in Eqs.~(\ref{CSS},\ref{sudakov}) have perturbative
expansions in $\alpha_s$,
\begin{eqnarray}
\label{aexp}
A_c(\as) &=& \sum_{n=1}^\infty \left( \frac{\as}{2\pi} \right)^n A_c^{(n)} \;\;, \\
\label{bexp}
B_c(\as) &= &\sum_{n=1}^\infty \left( \frac{\as}{2\pi} \right)^n B_c^{(n)}
\;\;, \\
\label{cexp}
C_{ab}(\as,z) &=& \delta_{ab} \,\delta(1-z) + 
\sum_{n=1}^\infty \left( \frac{\as}{2\pi} \right)^n C_{ab}^{(n)}(z) \;\;.
\end{eqnarray}

In order to obtain the coefficients
in Eqs.(\ref{aexp}-\ref{bexp}) at a given order, the differential cross-section at small $q_T$ has to be computed at the same order.
A comparison with the power expansion in
$\alpha_s$ of the resummed result in Eq.~(\ref{CSS}) allows
to extract the coefficients that
control the resummation of the large logarithmic terms.

In this letter we study the behaviour of cross-sections at small transverse momenta at
second order in $\as$ both in the quark and  gluon channels.
 We find
that the analytic form of the logarithmically-enhanced contributions can be
computed perturbatively in a universal manner by using the recent knowledge on the infrared
behaviour of tree-level \cite{BCM,tree} and one-loop \cite{1loop} QCD amplitudes. 
In this way, we are able to extract the coefficients $A_{c}^{(1)}$,
$B_{c}^{(1)}$, $C_{ab}^{(1)}$, $A_{c}^{(2)}$ and $B_{c}^{(2)}$ for {\it any}
$q\bar{q}$ or $gg$ initiated process in the class (\ref{class}).
Details on our calculation will be given
elsewhere \cite{paper}. Here we
only present and discuss our main results.

By following Ref.\cite{ds} we multiply the differential cross-section, calculated at parton level, by
$q_T^2$ and take moments with respect to $z=Q^2/s$ defining the dimensionless quantity:
\begin{equation}
\label{sigman}
\Sigma(N)=\int dz\, z^N \f{q_T^2 Q^2}{d\sigma_0/d\phi}\f{d\sigma}{dq_T^2
  dQ^2d\phi}\, .
\end{equation}
In the quark channel, for the sake of simplicity and in order to compare
our result for $\Sigma(N)$ to the one originally obtained for
Drell-Yan in Ref.\cite{ds}, we restrict our attention to the
 {\it  non-singlet} contribution to the cross-section defined by
\begin{equation}
\sigma^{NS}= \sum_{ff'} \left( \sigma_{q_f \bar{q}_{f'}} - \sigma_{q_f q_{f'}}
\right)\, .
\end{equation}

To have $q_T\neq 0$ at least one gluon must be emitted, thus $\Sigma(N)$ has
the expansion:
\begin{equation}
\Sigma(N)=\f{\as}{2\pi}\;\Sigma^{(1)}(N)+\left(\f{\as}{2\pi}\right)^2\Sigma^{(2)}(N)+\cdots
\end{equation}
In the following we will systematically neglect in $\Sigma(N)$ all
contributions that vanish as $q_T\to 0$.

In order to compute the small $q_T$ behaviour of $\Sigma(N)$ our strategy is
as follows. The singular behaviour at small $q_T$ is dictated by the infrared
(soft and collinear) structure of the relevant QCD matrix elements. At ${\cal O}(\as)$ this structure has been 
known for long time \cite{BCM}. Recently,
the universal functions that control the soft and collinear singularities
of tree-level and one-loop QCD amplitudes at ${\cal O}(\as^2)$
have been computed \cite{tree,1loop}.

By using this knowledge, and
exploiting the simple kinematics of the leading order subprocess, we were able
to construct {\em improved} factorization formulae that allow to control {\em
  all} infrared singular regions avoiding any problem of double counting \cite{paper}. We have
used these improved formulae to approximate the relevant matrix elements and 
compute the small $q_T$ behaviour of $\Sigma(N)$ in a completely universal manner.

The calculation at ${\cal O}(\as)$ is straightforward and we recover the
well-known results:
\begin{equation}
\label{sigmaq1}
\Sigma_{q{\bar q}}^{(1)}(N)=2C_F\,\log \f{Q^2}{q_T^2}-3C_F+2\gamma_{qq}^{(1)}(N)
\end{equation}
and
\begin{equation}
\label{sigmag1}
\Sigma_{gg}^{(1)}(N)=2C_A\log\f{Q^2}{q_T^2}-2\beta_0+2\gamma_{gg}^{(1)}(N) \, .
\end{equation}
 
Here $C_F=\f{N_c^2-1}{2N_c}$, $C_A=N_c$ and $T_R=1/2$ are the $SU(N_c)$ QCD colour factors,
$\beta_0=\f{11}{6}C_A-\f{2}{3} n_f T_R$ and $\gamma_{qq}^{(1)}(N)$,
$\gamma_{gg}^{(1)}(N)$ are the
quark and gluon
one-loop anomalous dimensions, respectively.
From Eqs.~(\ref{sigmaq1},\ref{sigmag1}) one obtains:
\begin{equation}
A_a^{(1)}=2 C_a~~~~~~~~~~~B_a^{(1)}=-2 \gamma_a~~~~~~~~~~~a=q,g
\end{equation}
where  $C_a$ and $\gamma_a$ are the coefficients of the leading $(1-z)^{-1}$
singularity and $\delta(1-z)$ term in the
one-loop
Altarelli-Parisi kernels $P_{aa}$, respectively,
\begin{equation}
C_q=C_F~~~~~~~~~~~C_g=C_A~~~~~~~~~~~\gamma_q=\f{3}{2}
C_F~~~~~~~~~~~\gamma_g=\beta_0\, .
\end{equation}

At this order it is possible to obtain also the coefficient $C^{(1)}_{ab}$ by considering the
$q_T$ integrated distribution and including the
renormalized virtual correction to the $LO$ amplitude $c{\bar c}
\to A_1+A_2\dots A_n$, summed over spins and colours, which, at ${\cal
  O}(\ep^0)$, can be written as\footnote{All our results are
  obtained using the factorization and renormalization prescriptions of the
  $\overline{MS}$ scheme and within the framework of conventional dimensional regularization.}
\begin{equation}
\label{1loop}
\cm^{(0)\dagger}_{c\bar{c}}(\phi)\cm^{(1)}_{c\bar{c}}(\phi)+{\rm
  c.c.}=\f{\as}{2\pi}\left(\f{4\pi\mu^2}{Q^2}\right)^\ep\f{\Gamma(1-\ep)}
{\Gamma(1-2\ep)}\left(-\f{2 C_c}{\ep^2}-\f{2 \gamma_c}{\ep}+{\cal A}_c(\phi)\right)|\cm^{(0)}_{c\bar{c}}(\phi)|^2.
\end{equation}
In Eq. (\ref{1loop}) the structure of the poles in $\ep=(4-d)/2$ is universal \cite{poli}
and fixed by the flavour of the incoming partons. The {\em finite} part ${\cal
  A}$ (which can  depend on
the kinematics of the final state non-colored particles) depends instead on
the particular process in the class (\ref{class}) we want to consider.
In the case of Drell-Yan  we have \cite{dy}:
\begin{equation}
\label{dy}
{\cal A}_q^{DY}=C_F\left(-8+\f{2}{3}\pi^2\right)\, ,
\end{equation}
whereas for Higgs production in the $m_{top}\to\infty$ limit the finite contribution
is \cite{higgs}:
\begin{equation}
\label{higgs}
{\cal A}_g^{H}= 5 C_A +\f{2}{3} C_A \pi^2 -3 C_F \equiv 11+2 \pi^2\, .
\end{equation}

By using the information in Eq.~(\ref{1loop}) we obtain for $C^{(1)}_{ab}$
\begin{eqnarray}
\label{coeff}
C^{(1)}_{ab} (z)= - {\hat P}^{\ep}_{ab}(z) + \delta_{ab}\, \delta(1-z) \left( C_a\,
  \f{\pi^2}{6}+\f{1}{2} {\cal A}_a(\phi) \right)
\end{eqnarray}
where ${\hat P}^{\ep}_{ab}(z)$ is
the ${\cal O}(\ep)$ term in the Altarelli-Parisi
${\hat P}_{ab}(z,\ep)$ splitting kernel, given by:
\begin{eqnarray}
{\hat P}_{qq}^\ep(z)&=&- C_F\, (1-z) \nn \\
{\hat P}_{gq}^\ep(z)&=&- C_F\, z \nn \\
{\hat P}_{qg}^\ep(z)&=&-2 T_R\,  z (1-z) \nn \\
{\hat P}_{gg}^\ep(z)&=& 0\, .
\end{eqnarray}

At order $\as$  the coefficients $A_a^{(1)}$ and $B_a^{(1)}$ are fully
determined by the {\it universal} Altarelli-Parisi splitting functions.
The function $C_{ab}^{(1)}$ depends instead on the process through
the one-loop corrections to the $LO$ matrix element.
The general expression in Eq.~(\ref{coeff}) reproduces correctly the
coefficient $C_{ab}^{(1)}$ for Drell-Yan \cite{ds}, Higgs production in the
$m_{top}\to\infty$ limit \cite{kauff} and $\gamma\gamma$ production
 \cite{2gamma}\footnote{The actual expression of the coefficient
$C^{(1)}_{qq}$ for $ZZ$ production reported in Ref.~\cite{ZZ} is not fully correct.}.

At second order in $\as$, two different contributions to
$\Sigma^{(2)}(N)$ have to be considered: the real correction corresponding to the emission of one
extra parton (i.e., two gluons or a $q\bar{q}$ pair) with respect to the
${\cal O}(\as)$ contribution, and its corresponding virtual correction.

The double-real emission contribution is the most difficult to compute. One has to integrate over the phase space of the two unresolved final state partons keeping $q_T$ fixed 
and finally perform the $z$ integration in Eq. (\ref{sigman}). We find that,
likewise $\Sigma^{(1)}(N)$, this contribution to $\Sigma^{(2)}(N)$ is {\em
  process independent}, i.e., it does not depend on the particular process in the class (\ref{class}) we want to consider.

The virtual contribution is simpler to compute and  we find it to be  {\em
  process dependent}. More importantly, its process dependence is fully
determined by the function ${\cal A}$ appearing in the one-loop correction
to the $LO$ subprocess (see Eq.~(\ref{1loop})).

In the following, for the sake of simplicity, we present the total results for
$\Sigma^{(2)}(N)$ corresponding to the choice of the factorization and
renormalization scales fixed to $Q^2$. Since we are interested in extracting
the coefficients $A^{(2)}_{q,g}$ and $B^{(2)}_{q,g}$, as in the ${\cal O}(\as)$ case we 
concentrate on the {\it diagonal} $q\bar{q}$ and $gg$ contributions  to
$\Sigma^{(2)}(N)$.

In the quark (non-singlet) channel we obtain: 
\begin{eqnarray}
\label{sigmanq}
\Sigma^{(2)}_{q\bar{q}}(N) &=& \log^3\f{Q^2}{q_T^2} \left[-2 C_F^2 \right] \nn\\
&+& \log^2\f{Q^2}{q_T^2} \left[ 9 C_F^2 +2 C_F \beta_0 -6 C_F \gamma_{qq}^{(1)}(N) \right] \nn\\
&+& \log\f{Q^2}{q_T^2} \left[ C_F^2 \left( \f{2}{3} \pi^2 -7  \right)
+ C_F C_A \left( \f{35}{18}-\f{\pi^2}{3} \right) 
-\f{2}{9} C_F n_f T_R  +2 C_F {\cal A}_{q}(\phi)
 \right. \nn\\
 & &  \hspace{1.2cm} \left.
+ \left(2 \beta_0 + 12 C_F\right) \gamma_{qq}^{(1)}(N) -4 \left(\gamma_{qq}^{(1)}(N)
\right)^2
 + 4 C_F^2 \left( \pen-\f{1}{2} \right) 
 \right] \nn\\
&+& \left[ C_F^2 \left( -\f{15}{4}-4 \zeta(3)   \right)
      + C_F C_A \left(-\f{13}{4} -\f{11}{18} \pi^2  +  6 \zeta(3)  \right)
-3 C_F {\cal A}_q(\phi)
 \right. \nn\\
 & & \left.
      + C_F n_f T_R \left( 1 + \f{2}{9} \pi^2  \right)
+ 2 \gamma_{(-)}^{(2)}(N) + 2 C_F \gamma_{qq}^{(1)}(N) \left( 
    \f{\pi^2}{3} +2 \pen
 \right) 
 \right. \nn\\
 & &  \hspace{0.0cm} \left.
+2 \gamma_{qq}^{(1)}(N) {\cal A}_q(\phi) 
 -2 C_F (\beta_0 +3 C_F) \left( \pen-\f{1}{2} \right)
    \right]\, ,
\end{eqnarray}
whereas for the gluon channel the result is:

\begin{eqnarray}
\label{sigmang}
\Sigma^{(2)}_{gg}(N) &=& \log^3\f{Q^2}{q_T^2} \left[-2 C_A^2 \right] \nn\\
&+& \log^2\f{Q^2}{q_T^2} \left[ 8 C_A \beta_0 -6 C_A \gamma_{gg}^{(1)}(N) \right] \nn\\
&+& \log\f{Q^2}{q_T^2} \left[ C_A^2 \left( \f{67}{9}+ \f{\pi^2}{3} \right)
-\f{20}{9} C_A n_f T_R   + 2 C_A {\cal A}_{g}(\phi)
 \right. \nn\\
 & &  \hspace{1.2cm} \left.
+ 2 \beta_0  \left( \gamma_{gg}^{(1)}(N)-\beta_0\right) -4 \left(
    \gamma_{gg}^{(1)}(N) -\beta_0\right)^2
-4 n_f\,\gamma_{gq}^{(1)}(N) \gamma_{qg}^{(1)}(N)
 \right] \nn\\
&+& \left[ C_A^2 \left( -\f{16}{3}+ 2 \zeta(3)\right) 
 + 2 C_F n_f T_R + \f{8}{3} C_A n_f T_R
-2 \beta_0 \left( {\cal A}_g(\phi) + C_A \f{\pi^2}{6} \right)
 \right. \nn \\     
& & \left. + 2 \gamma_{gg}^{(2)}(N) + 2 \gamma_{gg}^{(1)}(N) \left(  {\cal A}_g(\phi) +
    C_A \f{\pi^2}{3} \right) +4 C_F n_f\,\gamma_{qg}^{(1)}(N) \peng
    \right]\, .
\end{eqnarray}

In Eq.~(\ref{sigmanq}) $\gamma_{(-)}^{(2)}(N)$ is the non singlet space-like
two-loop anomalous dimension \cite{2loopns}, in Eq.~(\ref{sigmang}) $\gamma^{(2)}_{gg}(N)$ is the
singlet space-like two-loop anomalous dimension \cite{2loops} , 
$\zeta(n)$ is the Riemann $\zeta$ function ($\zeta(3)=1.202...$) and the function ${\cal
  A}_a(\phi)$ is defined through Eq.~(\ref{1loop}). The coefficients $\pen$
and $\peng$ have origin on the $N$ moments of  $-\hat{P}_{qq}^\ep(z)$ and
$-\hat{P}_{gq}^\ep(z)$, respectively.

The $N$-dependent part of the results in Eqs.~(\ref{sigmanq},\ref{sigmang})
agrees\footnote{This can also be regarded as an independent re-evaluation of the
  two-loop anomalous dimensions.} with the one obtained from the second order
expansion of Eq.~(\ref{CSS}) (see e.g. Ref.~\cite{AK}). By comparing also
the N-independent part we obtain for $A^{(2)}$:
\begin{equation}
A_a^{(2)}= K A_a^{(1)}~~~~~~~~~~~K=C_A \left( \f{67}{18}- \f{\pi^2}{6} \right) - n_f T_R \f{10}{9}
\end{equation}
in agreement with the results of Ref.~\cite{KT,cdet}.
Moreover we find that $B^{(2)}$ can be expressed as:
\begin{equation}
\label{b2}
B_a^{(2)}=-2\, \delta P_{aa}^{(2)} + \beta_0
\left( \f{2}{3} C_a \pi^2 + {\cal A}_a(\phi)  \right)~~~~~~~~~~~a=q,g
\end{equation}
where 
$\delta P_{aa}^{(2)}$ are the coefficients of the $\delta(1-z)$ term in the
two-loop splitting functions $P_{aa}^{(2)}(z)$ \cite{2loopns,2loops}, and are given by
\begin{eqnarray}
\delta P_{qq}^{(2)}&=& C_F^2\left( \f{3}{8}-\f{\pi^2}{2}+6 \zeta(3) \right) +
 C_F C_A \left( \f{17}{24}+ \f{11 \pi^2}{18}-3 \zeta(3)\right) 
- C_F n_f T_R\left( \f{1}{6}+ \f{2 \pi^2}{9}\right) 
\nn \\ 
\delta P_{gg}^{(2)}&=& C_A^2\left( \f{8}{3}+ 3 \zeta(3) \right) - C_F n_f T_R -
\f{4}{3} C_A n_f T_R\, .
\end{eqnarray}

From Eq.~(\ref{b2}) we see that $B^{(2)}$, besides
the  $-2\, \delta P_{aa}^{(2)}$ term which matches the expectation from the
${\cal O}(\as)$ result, receives a {\it process-dependent} contribution controlled
by the one-loop correction to the $LO$ amplitude (see Eq.~(\ref{1loop})).
We conclude that the Sudakov form factor in Eq.~(\ref{sudakov}) is actually process
dependent beyond next-to-leading logarithmic accuracy.
The interpretation of this result will be given elsewhere \cite{paper2}.

However, by using the general expression in Eq.~(\ref{b2}) it is possible to obtain
$B^{(2)}$ for a given process just by computing the one-loop correction to the
$LO$ amplitude for that process.
For the case of Drell-Yan, by using Eq.~(\ref{dy}), our
result for $\Sigma^{(2)}_{q\bar{q}}(N)$ agrees with the one of Ref.~\cite{ds}, 
and we confirm:
\begin{equation}
\label{b2dy}
B_q^{(2)DY}=C_F^2\left(\pi^2-\f{3}{4}-12\zeta(3)\right)+C_F\,C_A\left(\f{11}{9}\pi^2-\f{193}{12}+6\zeta(3)\right)+C_F\,n_f\,T_R\left(\f{17}{3}-\f{4}{9}\pi^2\right)\, 
.
\end{equation}
In the case of Higgs production in the $m_{top}\to\infty$ limit, by using
Eq.~(\ref{higgs}) we find:
\begin{equation}
\label{b2h}
B_g^{(2)H}=C_A^2\left(\f{23}{6}+\f{22}{9}\pi^2-6\zeta(3)\right)+4C_F\,n_f\,T_R-C_A\,n_f\,T_R\left(\f{2}{3}+\f{8}{9} \pi^2\right) 
 -\f{11}{2} C_F\, C_A \, .
\end{equation}

In particular, this result allows to improve the present accuracy of the matching
between resummed predictions \cite{resum} and fixed order calculations
\cite{higgsjet}.

Summarizing, we have studied the
logarithmically-enhanced contributions at small transverse momentum in
hadronic collisions at second
order in perturbative QCD. The calculation
was performed in an process-independent manner, allowing us to
show that the Sudakov form factor is actually process dependent beyond
next-to-leading logarithmic accuracy.
We have provided
a general expression for the coefficient $B^{(2)}$ for both quark and 
gluon initiated processes.

We would like to thank Stefano Catani and Zoltan Kunszt for many valuable 
discussions and comments, Christine Davies for providing us with a copy of her
Ph.D. thesis and Luca Trentadue, James Stirling and Werner Vogelsang for discussions.

\end{document}